\documentclass[a4paper,11pt]{article}
\usepackage{jinstpub} 
\usepackage{lineno}

\proceeding{Talk presented at the International Workshop on Future Linear Colliders (LCWS 2023), 15-19 May 2023. C23-05-15.3.}

\title{\boldmath The alignment of the C$^{3}$ Accelerator Structures with the Rasnik alignment system}

\author[a,b,1]{Harry van der Graaf\note{retired; corresponding author}}
\author[a]{, Niels van Bakel}
\author[c]{, Bram Bouwens}
\author[d]{, Martin Breidenbach}
\author[d]{, Andrew Haase}
\author[e]{, Joris van Heijningen}
\author[f]{, Anoop Nagesh Koushik}
\author[d]{, Emilio Nanni}
\author[a,g]{, Tristan du Pree}
\author[f]{, Nick van Remortel}
\author[d]{, Caterina Vernieri} 

\affiliation[a]{Nikhef, Science Park $105$, Amsterdam, The Netherlands}
\affiliation[b]{Delft University of Technology, Mekelweg $1$, Delft, The Netherlands}
\affiliation[c]{Amsterdam Scientific Instruments ASI, Science Park 106, Amsterdam, The Netherlands}
\affiliation[d]{SLAC, Stanford, CA, USA}
\affiliation[e]{Centre for Cosmology, Particle Physics and Phenomenology (CP3), Universit\'{e} catholique de Louvain,\\Louvain-la-Neuve, Belgium}
\affiliation[f]{University of Antwerp}
\affiliation[g]{Twente University, The Netherlands}

\emailAdd{vdgraaf@nikhef.nl}

\abstract{The Rasnik 3-point alignment system, widely applied in particle physics experiments and in the instrumentation of 
gravitational wave experiments, can be used as N-point alignment system by 'leap frog' N individual 3-point systems. The conceptual implementation of Rasnik chains in C$^{3}$ is presented. Then, the proper operation of a laser diode and a CMOS image sensor in liquid nitrogen has been verified. Finally, next plans for testing a small but complete system, immersed in liquid nitrogen, are presented.}


\begin{document}
\maketitle
\flushbottom

\section{The Rasnik 3-point alignment system}
\label{sec:intro}

The first \textbf{R}ed \textbf{A}lignment \textbf{S}ystem \textbf{Nik}hef (Rasnik) was developed in 1983 for the alignment of the muon chambers in the Muon Spectrometer of the L3 experiment at CERN \cite{beker_2019}. As sensor, 4-quadrant photodiodes were used. In 1993, CMOS image sensors
became available, and Rasnik evolved into a system in which a back-illuminated coded mask is projected, by means of a positive singlet lens, onto the image sensor. Since 2005, some 8000 of these Rasnik system are flawlessly operational in the ATLAS experiment at CERN.
With these Rasnik systems, composed from low-cost, commercially available components, the 2D alignment of three points can be
measured with a spatial resolution of 1\,nm.
The best \textit{resolution power} of 7\,pm/$\surd$Hz was obtained with an image frame rate of 250\,Hz \cite{ultimate}.

\begin{figure}[htb]
\centering
\includegraphics[width=10 cm]{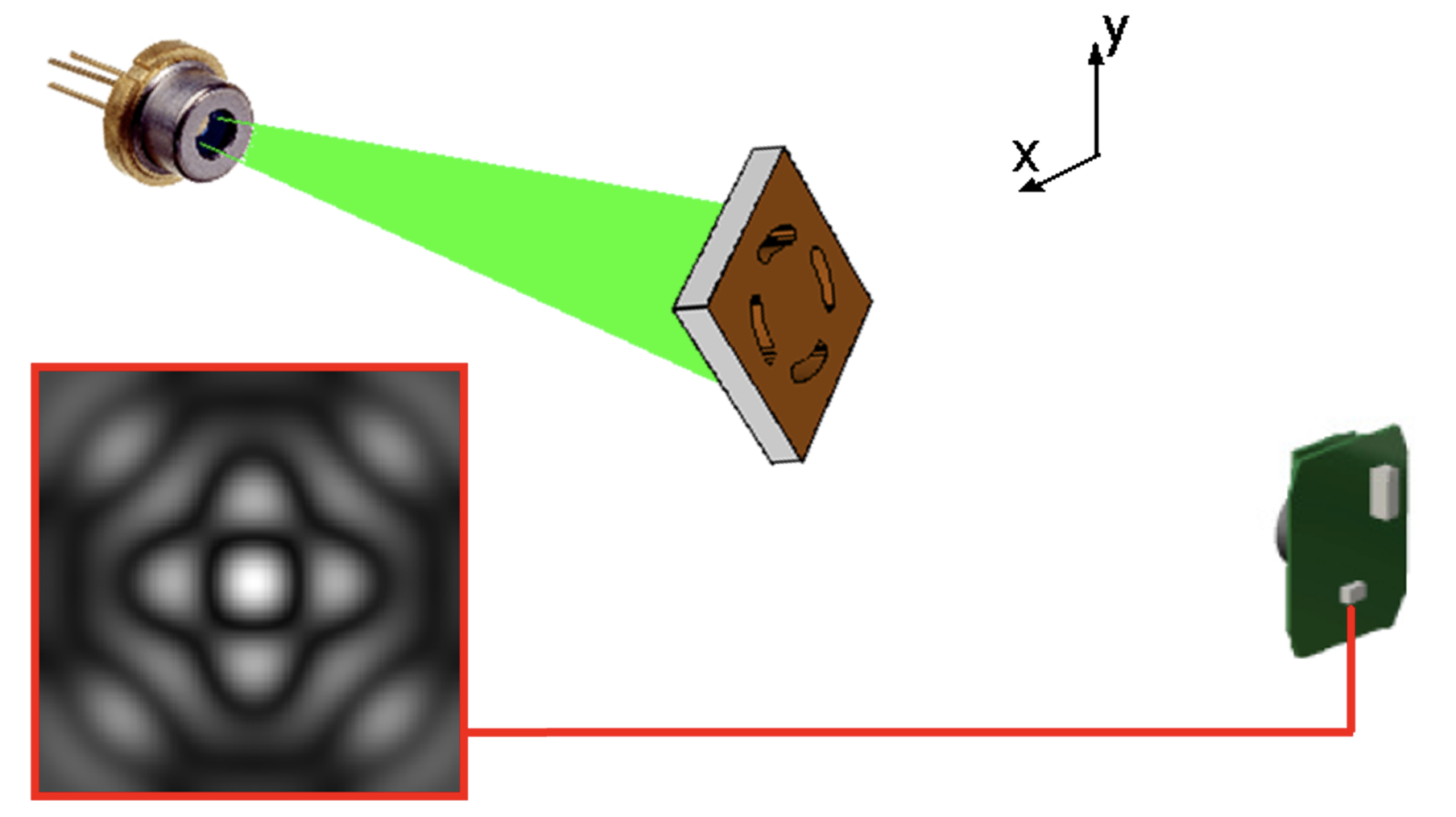}
\caption{Principle of the RasDif system. The monochromatic waves, arriving at the zone plate,
cause a diffraction pattern on the image sensor.
\label{fig:RasDif}}
\end{figure}

For the alignment of active beam elements of future CLIC and ILC linear colliders, systems with a span between the image sensor and the mask of a Rasnik system larger than 20 m are required. The diameter and focal length of the required lens become impractically large, and its replacement by a zone plate was considered. In the so-called RasDif system, depicted in figure \ref{fig:RasDif}, the back-illuminated coded mask is replaced by a monochromatic point-like light source generating spherical waves onto the zone plate. This results in a typical diffraction pattern onto the image sensor: the position of this pattern on the sensor is a measure for the 3-point alignment of light source, zone plate and image sensor, respectively.


\section{Rasnik in Liquid Nitrogen}

So far, Rasnik systems have been operated in ambient air and in vacuum. In C$^{3}$, Rasnik must operate in ambient air, in vacuum and in boiling liquid nitrogen (LN$_{2}$), at 79 K \cite{c3nr1} \cite{c3nr2}\cite{c3nr3}.
Given the index of refraction of LN$_{2}$ of 1.20, lenses can not be applied because a sharp image in LN$_{2}$ will be out of focus in vacuum. This is the main reason why the RasDif system is applied.

Laser diodes were tested to operate in LN$_{2}$: they work, albeit that their supply voltage of 2 V at 293 K usually needs to be raised to 6 V when immersed in LN$_{2}$. With a current of 10 mA, the heat dissipation causes local boiling and therefore the formation of bubbles. The glass cover seal provides local thermal shielding, so bubble formation in the critical light exit zone is locally reduced. The following three laser diodes were found to be LN$_{2}$ proof: Laser Components ADL65074TR, Laser Components ADL65055TL, and ROHM RDL65MZT7.
The long term stability of the sealed package in vacuum, in ambient air and in LN$_{2}$ remains to be verified.

Modern CMOS image sensors are known not to operate below 90 K. In 2014, two groups tested CMOS image sensors in populair low-cost
webcams by simply immersing them in LN$_{2}$, while reading them out via their USB connection\,\cite{McConkey, mavrokoridis}.
Eventually, two webcams were found to operate unaffected in LN$_{2}$: the Floureon "car reversing camera",
and the Microsoft HD-3000 Model 1456 (or earlier). The latter can still be acquired, and proper operation in LN$_{2}$ has been recently confirmed, including a cold start, avoiding the self heating of the circuitry due to dissipation. For the coming R\&D projects, some 8 of these image sensors will be applied in experimental set-ups.

It is well known, in CMOS design, which circuitry could be applied for a given constraint in temperature \cite{ghibaudo, hongliang, gong, patterson}. There is commercial interest in the development of a state-of-the-art CMOS image sensor
for monitoring the inside of LH$_{2}$ tanks.

\section{Rasnik in C$^{3}$}

\begin{figure}[htb]
\centering
\includegraphics[width=12 cm]{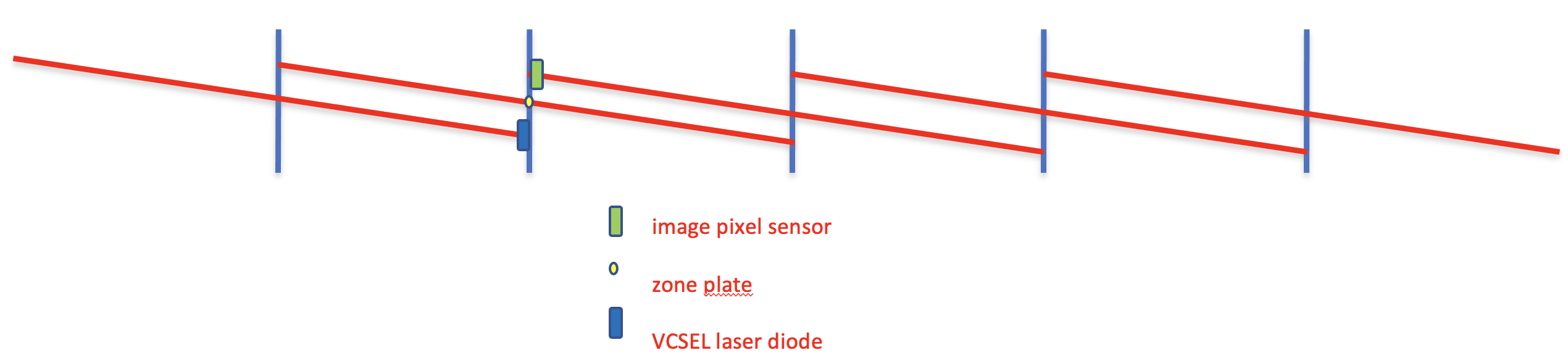}
\caption{Principle of the leapfrog multipoint alignment system. Each chain plate is equipped with a laser diode, a zone plate and an image sensor.}
\label{fig:leapfrog}
\end{figure}

In figure \ref{fig:leapfrog}, the assembly of chain plates is shown: the position, in X and Y, of any plate with respect to any other plate is known. The alignment of Accelerator Structures and Quads of C$^{3}$ can be realised by linking them with chain plates.

\begin{figure}[htb]
\centering
\includegraphics[width=12 cm]{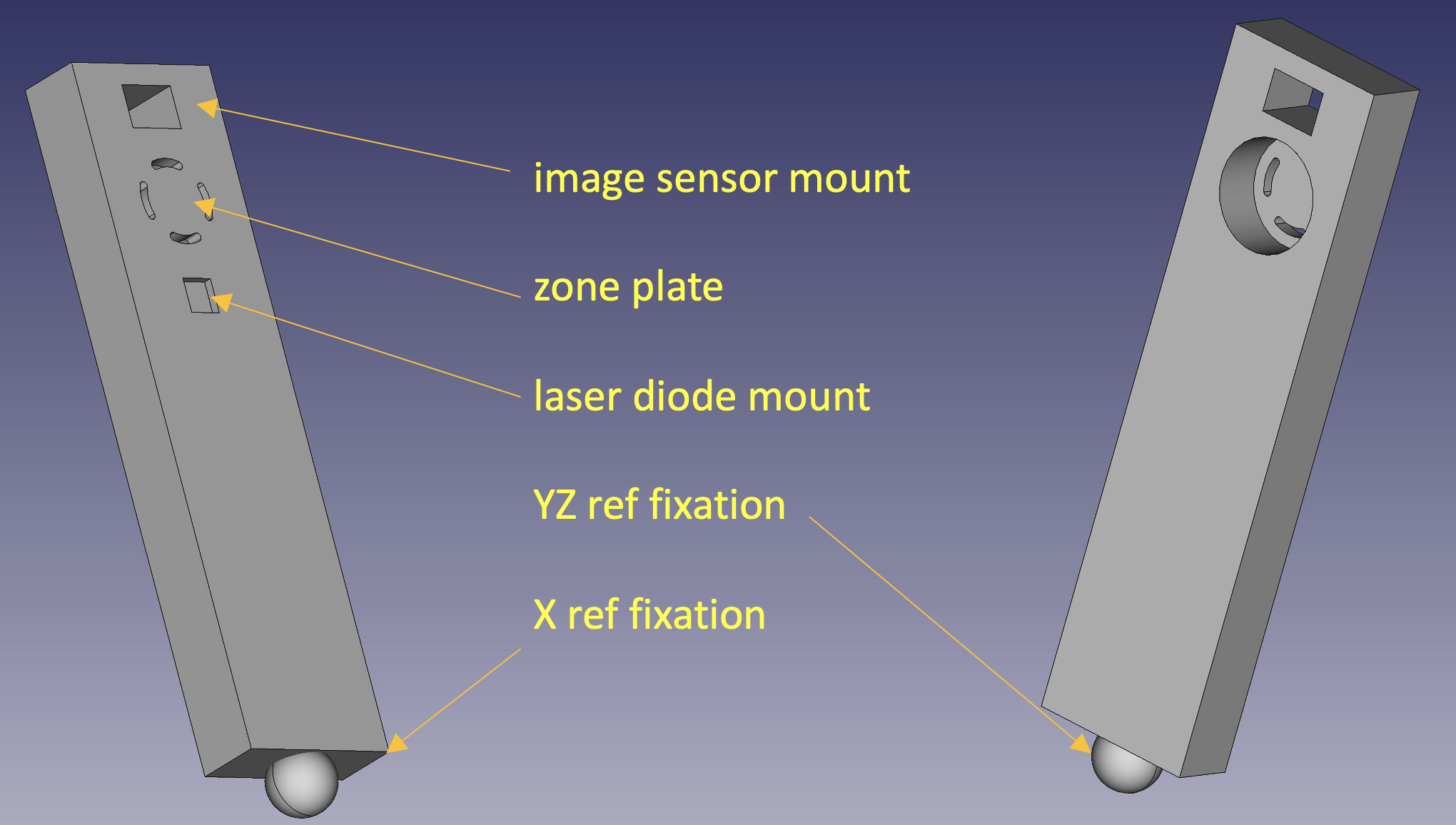}
\caption{The Stick. Each Stick includes the three basic Rasnik components. At one side, a Stick is fixed onto the object-to-align. It will be a challenge to reduce the associated 'contact error' to below 0.1 \textmu m.}
\label{fig:stick}
\end{figure}

In C$^{3}$, the chain plate takes the form of a stick, shown in figure \ref{fig:stick}. In a chain, all Sticks are practically identical: individual differences are known after a calibration procedure. On a Stick a CMOS image sensor chip is fixed, a pattern forming a zone lens is milled out, and laser diode is mounted. A Stick is mounted, using its mechanical
interface, onto an Accelerator Structure or a Quad, the latter being an assembly of a static quadrupole magnet and a Beam Position Monitor.

\begin{figure}[htb]
\centering
\includegraphics[width=12 cm]{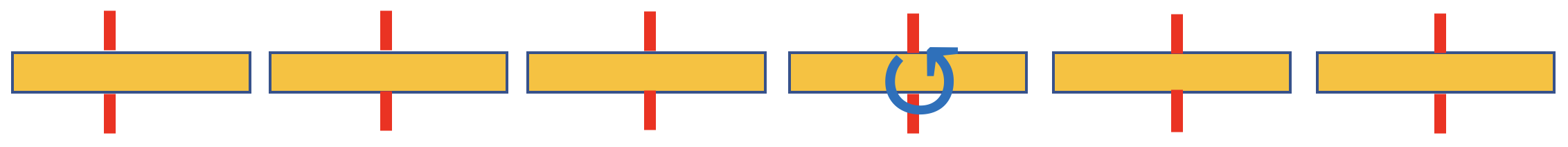}
\caption{Top view of the possible placement of Sticks onto the Accelerator Structures. Note that a rotation of a Structure around
its Y axis (perpendicular to the plane of view) is not recorded by Rasnik.}
\label{fig:basic}
\end{figure}

\begin{figure}[htb]
\centering
\includegraphics[width=12 cm]{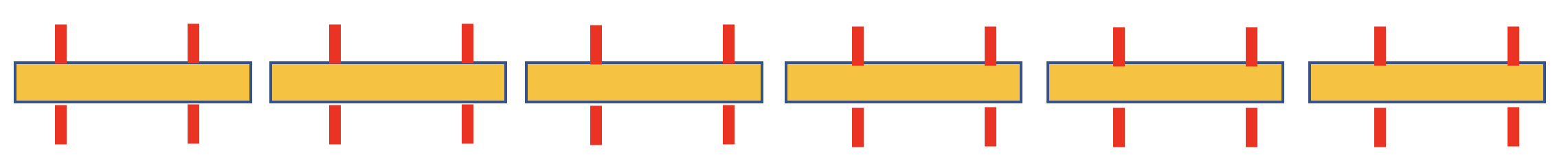}
\caption{Top view of the possible placement of Sticks onto the Accelerator Structures, providing complete and redundant data. The alignment of the Quads will be integrated in this chain.}
\label{fig:double}
\end{figure}

A first approach towards the application of Rasnik in the alignment of the 1 m long copper Accelerator Structures is shown in figure \ref{fig:basic}. Although there is some redundancy, the Rasnik data does not cover all 6 degrees of freedom of the Structures. Since two
adjacent Structures are confined by means of a \textit{raft} structure, this may be acceptable. In figure \ref{fig:double} the number
of Sticks is doubled, providing the redundant measurement of all 6 degrees of freedom of all Structures.

\section{The effective precision of a chain of Rasnik systems}
The error of a 3-point Rasnik system is defined as the variation in the \textit{sagitta}, defined as the distance, in X and Y,
of the optical centre of the zone plate to the optical axis, being the line through the optical centres of the laser source and the image sensor. Assuming that the medium is homogeneous, light propagates in a straight line. For simple systems using low cost laser diodes and image sensors extracted from popular webcams, the (Gaussian) noise in the sagitta measurement due to quantum fluctuations in light falling onto pixels, can be as good as 100 nm. With modern low-cost image sensors, the frame rate of 100 Hz results in a 
\textit{resolution power} of 10 nm/$\sqrt{Hz}$ which is the equivalent of 100 nm resolution of one measurement per second.
With a custom designed image sensor optimised for application in C$^3$, a resolution power of 5 nm/$\surd$Hz
should be within reach \cite{beker_2019, ultimate,JorisMSc}.

The 2D position of a Quad or Accelerator Structure is transferred to Rasnik by means of Sticks. The mechanical interface between the Stick and the object onto which it is mounted is associated with a \textit{set up error} of order 1 \textmu m if state-of-the-art
suspension mechanics are applied. In addition, the positions of
the optical centres of laser source, zone plate and image sensor with respect to the mechanical interface, which are measured in a calibration procedure (see \textbf{Calibration of the Sticks}) are subject to a similar \textit{offset error} of 1~\textmu m. 

Combining the two errors together, each Rasnik data point from a particular system is subject to an offset error of order 1 \textmu m, constant in time, and a random Gaussian error much smaller than 0.1 \textmu m.

\begin{figure}[htb]
\centering
\includegraphics[width=10 cm]{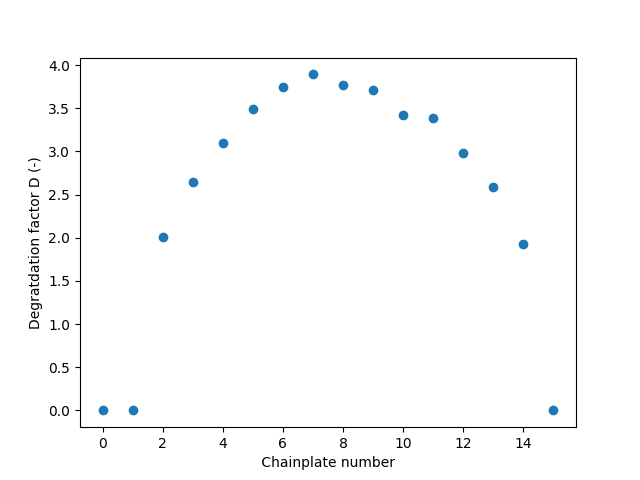}
\caption{The error propagation in a chain of N = 16 Rasnik systems.
The maximum error reaches D times Rasnik's intrinsic error.}
\label{fig:degra}
\end{figure}

Figure \ref{fig:degra} shows the errors associated with a chain of $N$ = 16 Rasnik systems. In this calculation the positions of chainplate 0, 1 and 15 are fixed, and the positions of the in-between chainplates are calculated from the MonteCarlo simulated Rasnik data, assuming a spatial resolution of 1.0 (RMS, unitless) for each system. The largest error occurs, as expected, at the central plate since it is far away from the fixed chainplates. The largest spatial error equals $\sqrt{N}$ times the sagitta error of a single Rasnik system, where D = $\sqrt{N}$.

The Quarter Cryogenic Module (QCM) will be equipped with 4-fold Rasnik chains placed at both sides. Given the offset error of
an individual Rasnik system of 1 \textmu m, and a degradation factor $\sqrt{N}$ = 2, the relative positions of the two Accelerator Structures and the Quad will be known within 1.5 \textmu m after the eight calibrated Sticks are placed.

In C$^3$, a SuperSector includes 728 Accelerator Structures. If these are equipped with two continuous Rasnik chains, the largest uncertainty in the middle of each chain is $\sqrt{728}$ / $\sqrt{2}$ = 19 \textmu m. By adding Rasnik systems on Sticks 1, 364 and
728, this largest error can be reduced by a factor $\sqrt{2}$. As long as the light path area, now confined by the raft's main beams, is respected, there is room for improvement.

\section{Conclusions and future plans}
\label{sec:conclusions}

\begin{figure}[htb]
\centering
\includegraphics[width=10 cm]{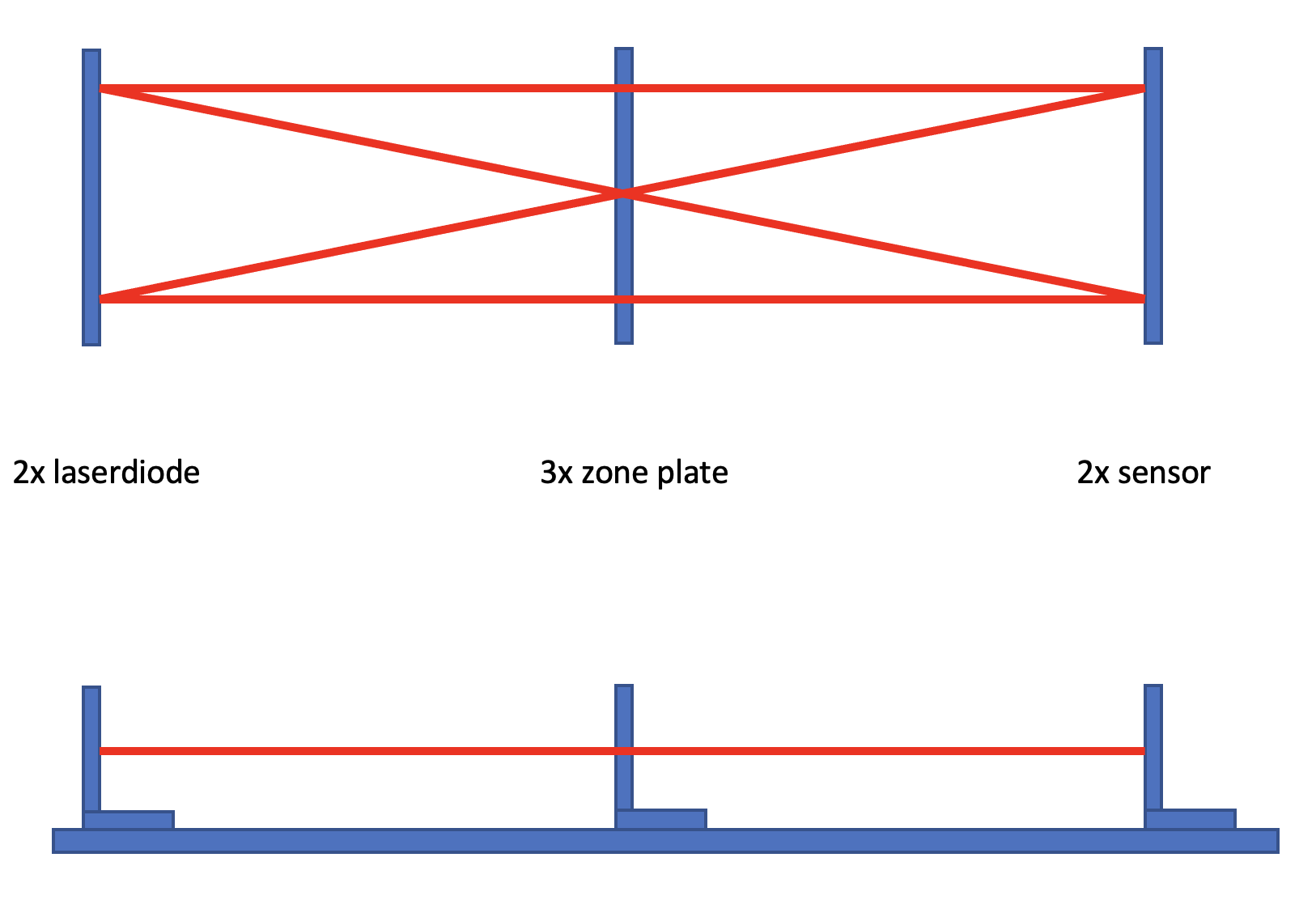}
\caption{The InPlaneDemo setup: it includes two mutually coupled parallel Rasnik systems, and two 'diagonal' systems. With this
2 to 4 fold redundant system, the quality of Rasnik systems can be verified in ambient air, vacuum,
and fully immersed in LN$_{2}$ \cite{beker_2019}. This setup fits in a foam cryostat with inner fiducial dimension of 60 x 10 x 10 cm$^{3}$}
\label{fig:InPlane}
\end{figure}

\begin{figure}[htb]
\centering
\includegraphics[width=12 cm]{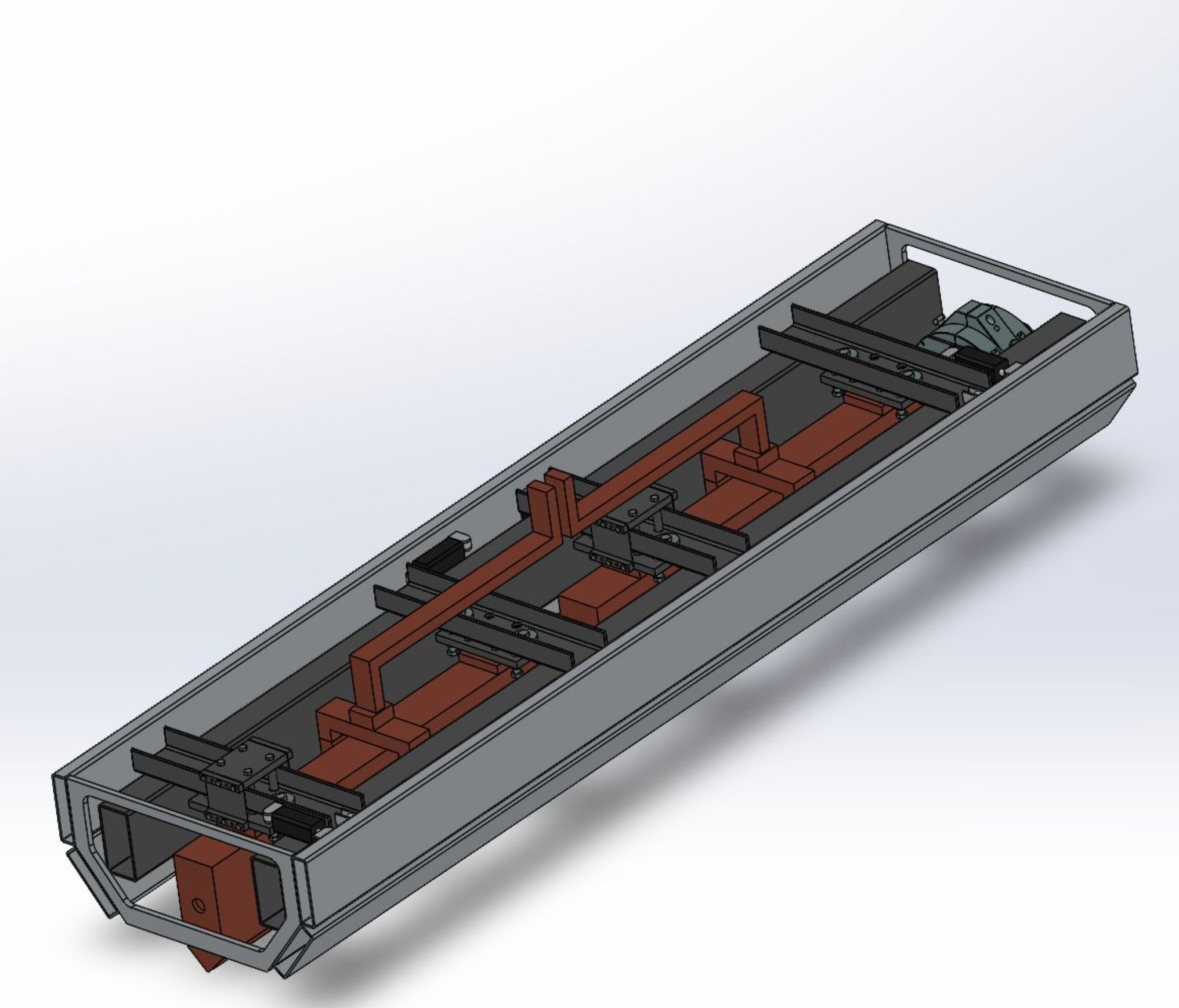}
\caption{The raft unit.}
\label{fig:oneraft}
\end{figure}

\begin{figure}[htb]
\centering
\includegraphics[width=10 cm]{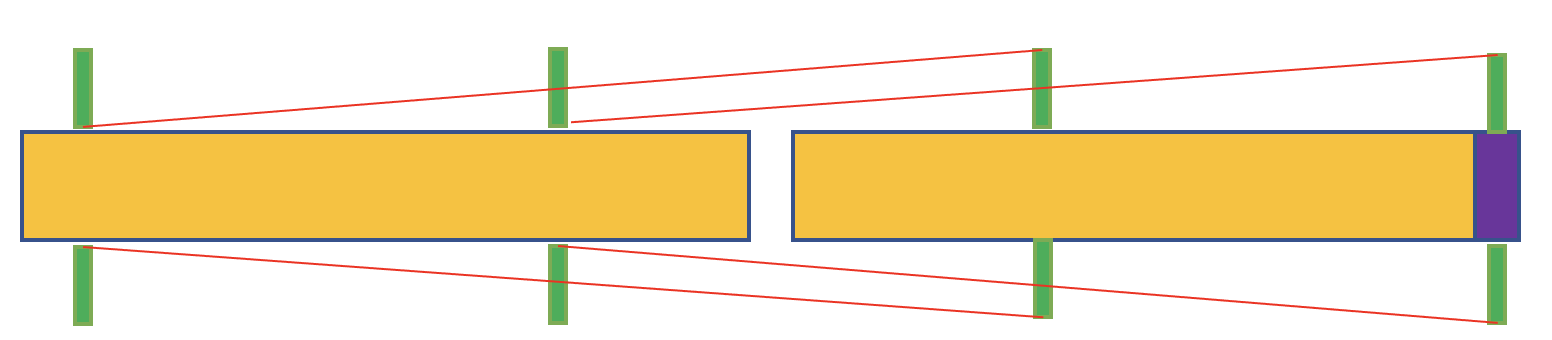}
\caption{Layout of the 8 Sticks in QCM. With calibrated Sticks, the relative positions of the two Accelerator Structures and the Quad are known.}
\label{fig:QCMsticks}
\end{figure}

\begin{figure}[htb]
\centering
\includegraphics[width=10 cm]{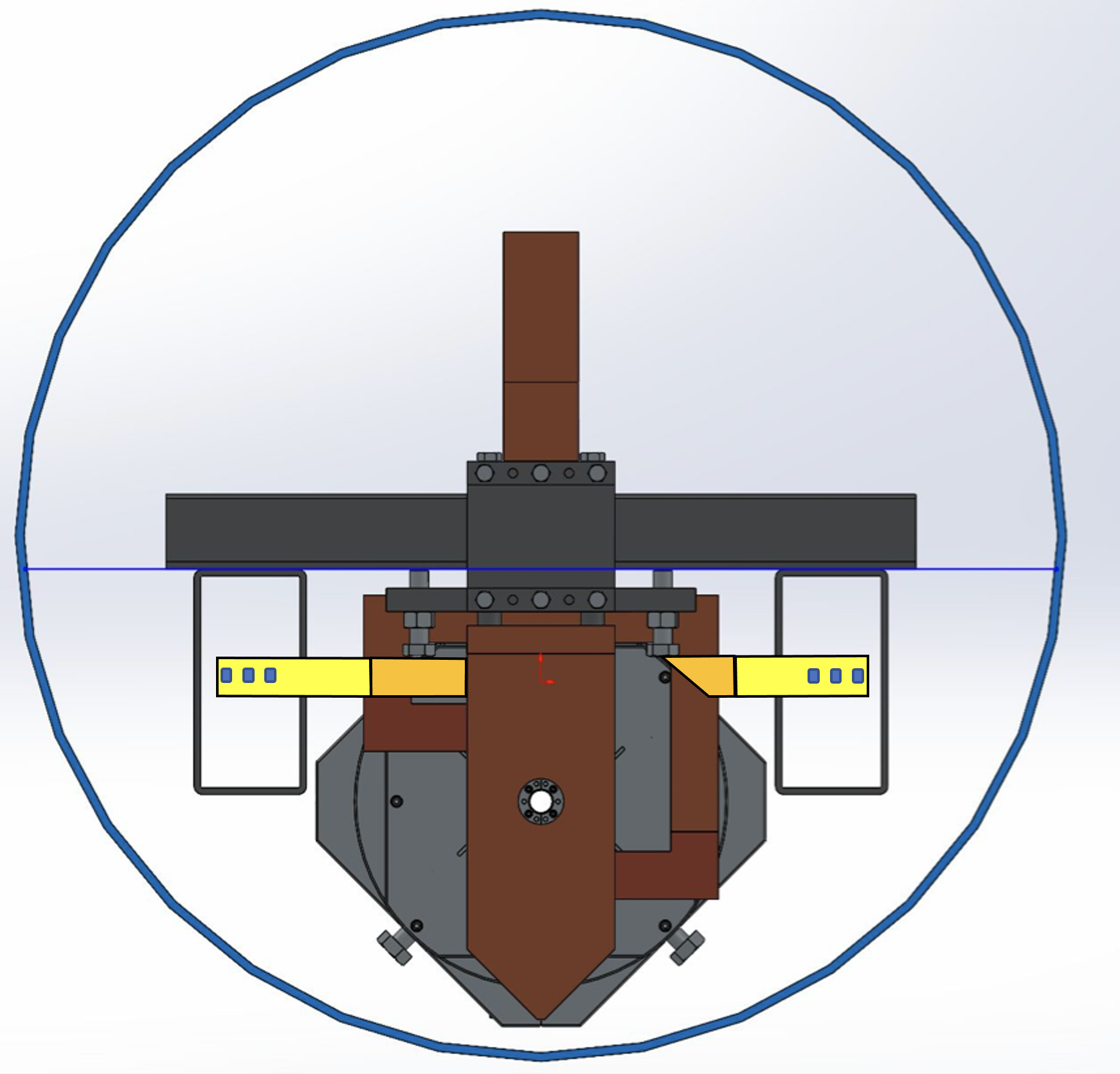}
\caption{Cross section of the QCM. Since all Sticks are identical, interfaces are required between a Stick and either an Accelerator Structure (left) or the Quad (right). These interfaces may be an integral part of the Accelerator Structure or the Quad.}
\label{fig:QCMcross}
\end{figure}

One CMOS image sensor has been found and identified to operate flawlessly immersed in LN$_{2}$ at 77~K. With some 10 of these sensors,
demonstrations of complete Rasnik systems, in air and immersed in LN$_{2}$, could be performed, within a year. Most likely, new CMOS image sensors will become available, capable to operate in a cryogenic environment, thanks to commercial interest. The proper operation of laser diodes in LN$_{2}$ has been certified.

By applying Rasnik instead of a system in which the local position of a stretched wire
is determined by multiple Wire Position Sensors (WPS) \cite{helene1},\cite{helene2}, severe problems are avoided: no risk of a hard-accessible broken wire,
no error in the vertical coordinate due to uncertainties in wire sag, and no risk of variations in the wire position due to flowing LN$_{2}$ medium.

Other benefits of using Rasnik alignment systems are:

\begin{itemize}
	\item electronics: benefits of using only off-the-shelf USB Muxs, CPUs or GPUs;
	\item precision is limited by (mechanical) offset; intrinsic 2D spatial resolution of 10 nm;
 	\item mechanical noise, due to bubble formation and flow of LN$_{2}$ can be well studied given the
          excellent spatial resolution power;
	\item no scale calibration required;
	\item low cost: € 30 per system, excluding readout;
	\item no drift: 1/f noise is not measurable;
 	\item radiation-proof components can be selected.
\end{itemize}

Figure \ref{fig:InPlane} shows the \textbf{InPlaneDemo} setup. It includes two parallel and two diagonal Rasnik systems.
With this, performance-limiting effects of Rasnik in LN$_{2}$ can be studied in detail: 

\begin{itemize}
	\item the formation of N$_{2}$ gas bubbles on the surface of the heat-dissipating laser diodes and CMOS image sensors;
	\item local variations in the index of refraction of LN$_{2}$ due to the laminar or turbulent flow, and due to convection near the warmer laser diodes and CMOS image sensors;
	\item mechanical noise due to bubble formation and bubble transport in LN$_{2}$. Here the 10 nm spatial resolution and a high frame rate of Rasnik will be useful;
	\item the long-term quality degradation of laser diodes and CMOS image sensors.
\end{itemize}

\textbf{Rasnik for the Quarter Cryogenic Module (QCM).} Figure \ref{fig:oneraft} shows the basic \textbf{raft} unit: an assembly of two
sequential Accelerator Structures, one Quad module, waveguides and a frame keeping these items in their proper position. Placed in a cryostat filled with LN$_{2}$, the functionality of QCM can be fully tested \cite{marty}. For this, the relative positions of
the Quad and the Accelerator Structures can be monitored with Rasnik, as is shown in figures \ref{fig:QCMsticks} and \ref{fig:QCMcross}.
The optical paths are confined in the raft's construction bars. Since only the laser diodes and image sensors dissipate their power there is little bubble formation inside these bars, and the LN$_{2}$ flow in the bar is expected to be directed in Z, and laminar. Variations in the index of refraction in the X and Y directions should be small.

\begin{figure}[htb]
\centering
\includegraphics[width=10 cm]{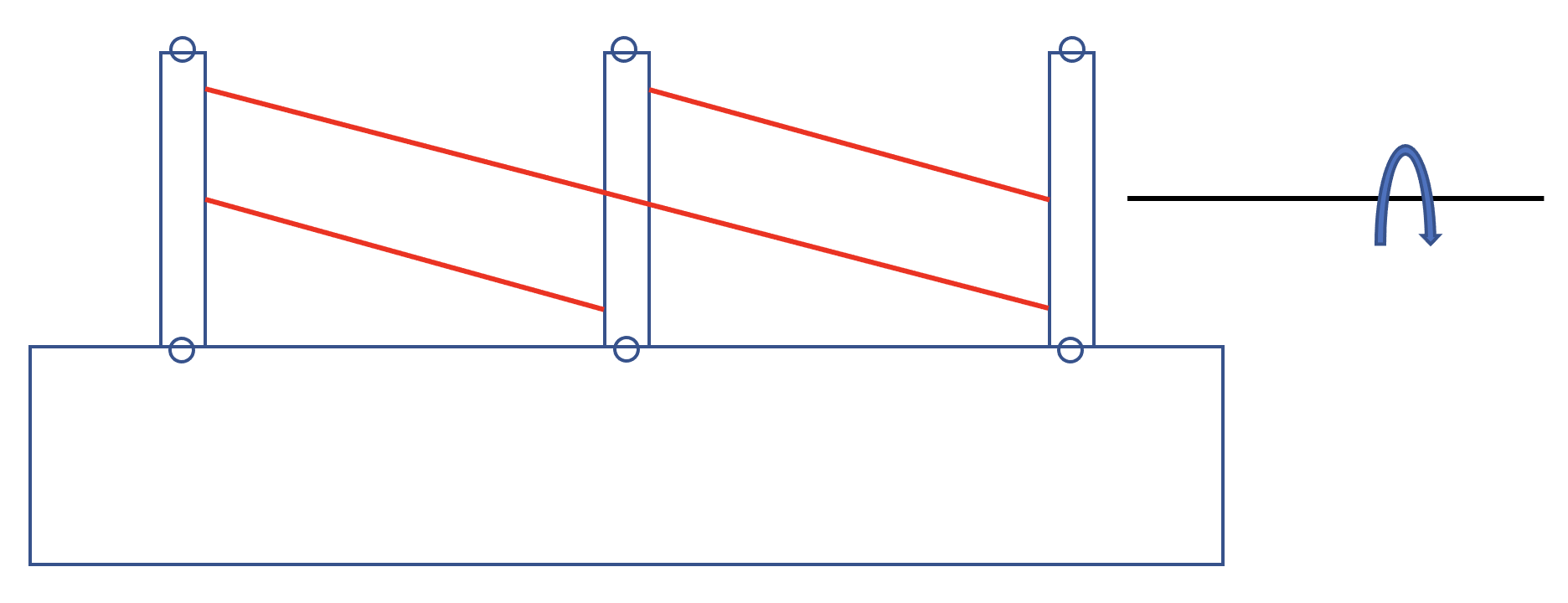}
\caption{The Calibration Station. Three special Sticks have their mechanical references at both ends, such that the distance
between the two references is precisely identical for all three Sticks. After rotating all three Sticks around the Z axis by 180 deg, the image shift on the sensor on the right-hand Stick equals 4x the \textbf{sagitta},
being the alignment error in X and Y of the Station.}
\label{fig:calib}
\end{figure}

\textbf{Calibration of the Sticks.} With calibrated Sticks, the 6D relative positions of the two Accelerator Structures and the Quad
in the QCM are known as soon as the  eight Sticks are mounted in position. With the \textbf{Calibration Station}, shown in figure \ref{fig:calib}, placed in a normal lab environment, the 3x X and 3x Y coordinates of the source, zone plate and sensor, respectively, with respect to the mechanical reference, can be obtained for each individual Stick \cite{atlas}. After measuring the alignment error of the Station, the calibration of any Stick is obtained by placing sets of three Sticks on the station, such that each Stick will pass at least one time the left-, middle- and right position on the Station.

\acknowledgments

We thank Nikhef for facilitating tests of crucial components in their dewar, and we are grateful to Amolf to
provide us with LN$_{2}$. We thank Oscar van Petten for producing mechanical supports, Berend Munneke for his practical assistance, and Nico Rem for keeping us working safe.

\end{document}